\documentclass[aps,prl]{revtex4}

\usepackage{amsmath,amssymb}
\def\gfxon{\usepackage[final]{graphicx}}

\gfxon

\usepackage[T1]{fontenc}
\usepackage{amsthm}
\usepackage{amsfonts}
\usepackage{bm}
\usepackage{cancel}
\usepackage[center]{subfigure}

\usepackage{booktabs}
\usepackage{xcolor}

\usepackage{soul}

\makeatletter
\let\old@startsection=\@startsection
\renewcommand{\@startsection}[6]{\old@startsection{#1}{#2}{#3}{#4}{#5}{#6\mathversion{bold}}}
\makeatother

\makeatletter
\def\mr@ignsp#1 {\ifx\:#1\@empty\else #1\expandafter\mr@ignsp\fi}%
\newcommand{\multiref}[1]{\begingroup%\let\protect\string%
\xdef\mr@no@sparg{\expandafter\mr@ignsp#1 \: }%
\def\mr@comma{}%
\@for\mr@refs:=\mr@no@sparg\do{\mr@comma\def\mr@comma{,}\ref{\mr@refs}}%
\endgroup}
\makeatother

\DeclareGraphicsExtensions{.pdf,.png,.jpg}

\ifx\href\asklfhas\newcommand{\href}[2]{#2}\fi

\begin{document}

\title{Crossover between integer and fractional vortex lattices\\ 
in coherently coupled 
two-component Bose-Einstein condensates}

\author{Mattia Cipriani$^{1,2}$, Muneto Nitta$^{3}$}

\affiliation{$^1$University of Pisa, Department of Physics ``E. Fermi'', Largo Bruno Pontecorvo 3, 56127, Italy \\
$^2$Istituto Nazionale di Fisica Nucleare - Sezione di Pisa, Largo Bruno Pontecorvo 3, 56127, Italy
 \\
$^3$Department of Physics, and Research and Education Center 
for Natural Sciences, Keio University, 4-1-1 Hiyoshi, Yokohama, 
Kanagawa 223-8521, Japan
}

\begin{abstract}
We study effects of the internal coherent (Rabi) coupling in vortex lattices
in two-component BECs under rotation. 
We find how the vortex lattices 
without the Rabi coupling known before are connected to 
the Abrikosov lattice of integer vortices with increasing the Rabi coupling. 
We find that 
1) for small Rabi couplings, fractional vortices in triangular or square lattice 
for small or large inter-component coupling
 constitute hexamers or tetramers, 
namely multi-dimer bound states made of six or four vortices, 
respectively,  
2) these bound states are broken into a set of dimers at  
intermediate Rabi couplings, 
and 3) vortices change their partners 
in various ways depending on the inter-component coupling
to organize themselves for 
constituting the Abrikosov lattice of integer vortices 
at strong Rabi couplings.
\end{abstract}

%{\footnotesize \texttt{IFUP-TH/2013-08 } }

\maketitle

\sloppy

Multi-component condensations are 
one of growing topics in condensed matter physics such as 
exotic superconductors, superfluid $^3$He, 
multi-component or spinor Bose-Einstein condensates (BECs) of ultra-cold atomic gases, 
exciton-polariton condensates, nonlinear optics, and nonlinear sciences. 
They also appear in high energy physics and astrophysics such as 
hadronic matter composed of neutron and proton Cooper pairs 
relevant for cores of neutron stars,  
and quark matter composed of di-quark condensates 
consisting of quark Cooper pairs  
which might be present at more high density. 
One of new common features of these systems is the existence of exotic vortices created by rotating superfluids or BECs or by applying magnetic field on superconductors. 
There exist different vortices winding around different condensates 
in general. 
Their quantized circulations for superfluids or BECs 
and fluxes for superconductors are not integer valued anymore
but are rational or fractional in general, 
as found in various systems: superfluid $^3$He \cite{Salomaa:1985,Volovik:2003}, 
p-wave superconductors \cite{Salomaa:1985,Ivanov:2001,Chun:2007,Jang:2011},   
multi-gap superconductors  \cite{Babaev:2002,Goryo:2007},
spinor BECs \cite{Ho:1998,Semenoff:2006vv}, 
multi-component BECs
\cite{Mueller:2002,Son:2001td,Kasamatsu:2003,Kasamatsu:2005,Kasamatsu:2004,Kasamatsu:2009,
Eto:2011wp,Aftalion:2012,Kuopanportti:2012,Eto:2012rc}, 
exciton-polariton condensates \cite{exciton,exciton-lattice}, 
nonlinear optics \cite{optics},
and color superconductors as quark matter \cite{Balachandran:2005ev}. 

Among various condensed matter systems admitting vortices, 
BECs of ultra-cold atomic gases provide a particularly 
ideal system for examining properties of vortices 
both theoretically and experimentally \cite{Pethick:2008}.  
Theoretically, BECs can be quantitatively well described in the mean-field theory, {\it i.e.}, the Gross-Pitaevskii (GP) equation.
On the other hand,  
BECs are quite flexible and controllable systems experimentally, 
because the atomic interaction is tunable through 
a Feshbach resonance \cite{Chin:2010}  
and the condensates can be visualized directly 
by optical techniques.  
Two-component BECs have been realized by using the mixture
of atoms with two hyperfine states of 
$^{87}$Rb \cite{Myatt:1997}
or the mixture of two different species of atoms 
\cite{Modugno:2002,Papp:2008,McCarron:2011}.

One of fascinating features of fractional vortices 
is the possibility of various structures of vortex lattices, 
as found in a vortex phase diagram 
in two-component BECs \cite{Mueller:2002,Kasamatsu:2003,Kasamatsu:2005,Aftalion:2012,
Kuopanportti:2012}.
When the inter-component coupling is increased, 
one obtains from an Abrikosov's triangular lattice 
of fractional vortices to 
a square lattice of fractional vortices \cite{Mueller:2002,Kasamatsu:2003,Kasamatsu:2005}, 
and a vortex sheet \cite{Kasamatsu:2009}. 
One may expect a similar phase diagram 
in exotic superconductors 
if the rotation speed is replaced with an applied
magnetic field.
However, for multi-gap superconductors, 
the existence of a Josephson coupling between 
different condensates is inevitable.  
While this term provides a gap to the Leggett mode 
corresponding to 
the phase difference between two condensates, 
it has been predicted 
{
in Refs.~\cite{Babaev:2002,Goryo:2007} 
}
that it also 
binds fractional vortices winding around two 
different components by a sine-Gordon kink {\cite{Tanaka:2001,Gurevich:2003}}, 
resulting in a two-vortex molecule, a {\itshape dimer}.
However, such a molecule structure has not yet been 
observed in exotic superconductors 
except for indirect evidences \cite{fractional-exp}.
In two-component BECs 
of atoms with two hyperfine states such as
$^{87}$Rb \cite{Myatt:1997}, 
two condensates can be coherently coupled 
by introducing a Rabi oscillation, 
which gives the same interaction as 
the Josephson coupling in superconductors. 
In fact, with the Rabi couplings, 
vortices winding around two (or more) different condensates 
are found to constitute a dimer  
\cite{Kasamatsu:2004,Kasamatsu:2005} 
(or a trimer \cite{Eto:2012rc}),  
connected by one (or more) sine-Gordon kink(s) \cite{Son:2001td}. 
{ 
Similar objects were discussed in spinor BECs~\cite{Turner:2009}.
}
The molecule structures are  
more accessible in BECs than superconductors
because of stronger repulsion between two kinds of
vortices and tunable atomic interactions in experiments in BECs.
However, effects on this term were not studied in 
the vortex phase diagram of two-component BECs.

The purpose of this Letter is to systematically study 
effects of the Rabi coupling in vortex lattices 
in two-component BECs under rotation. 
In the limit of a strong Rabi coupling, each vortex molecule 
is tightly bound to become an integer vortex, 
where one can expect the usual Abrikosov lattice of integer vortices. 
The effects of the coherent coupling were simulated in \cite{Woo:2008}
for an atomic-molecular BECs mixture.
However, the richer vortex lattice structure in atomic two-component BEC 
gives rise to a wider variety of configurations, 
as shown below. 
A quite nontrivial question is how the vortex lattices 
without the Rabi coupling known before are deformed into 
the integer Abrikosov lattice 
when the Rabi coupling is gradually increased. 
We find various new structures of vortex lattices 
and new phenomena of vortices, {\it i.e.},  
a bound state of multiple dimers  
bound by an intermolecular force 
and exchanging partners among multiple multi-dimer bound states  
analogous to chemical reactions.   
For small Rabi couplings, fractional vortices in triangular or square lattice constitute vortex hexamers or tetramers, respectively. 
When the Rabi coupling is increased, they are broken into a set of 
dimers. Then, they exchange their partners to organize themselves to prepare for becoming the Abrikosov lattice of integer vortices.

The energy functional of the Gross-Pitaevskii equations for the rotating BECs subject 
to a trapping harmonic potential can be written as:
\begin{align}\label{eq:FreeEnergy}
	E = & \int d^3 r \left\{ \sum_{\alpha=1,2} \Psi^*_\alpha \left[ \frac{1}{2} \left( \frac{1}{i} \bm{\nabla} - \Omega \, \hat{\bm{z}} \times \bm r \right)^2 + \frac{r^{2}}{2} \left( 1 - \Omega^2 \right) - \mu_\alpha \right] \Psi_\alpha + \frac12 g_1 |\Psi_1|^4 + \frac12 g_2 |\Psi_2|^4 + g_{12} |\Psi_1|^2 |\Psi_2|^2  \right\} 
\end{align}
where the derivatives include $(r,\theta)$ coordinates.  
We measure distances and energies in terms of $b_{\text{ho}} = \sqrt{\hbar/m\omega}$ and $\hbar \omega$ respectively, where $m$ is the mass of the atoms and $\omega$ is the frequency of the trapping harmonic potential.
{We consider $g_1=g_2=g$}.

The phase diagram of the vortex lattice forming in the condensate was studied in \cite{Mueller:2002,Kasamatsu:2003} and a rich variety of lattices was found.
The structure of the two component lattice depends on the sign and on the magnitude of $g_{12}$, the coupling constant of the intercomponent interactions. 

When $g_{12}/g=\delta<0$, the vortices of different components are attracted and are combined into integer vortices.
Because of repulsion among them, they organize in a triangular lattice. 
If $\delta>0$, depending on the value of $\delta$ and $\Omega$, vortices 
in each component organize in the triangular lattice, the square lattice, or the vortex-sheet.  
When $\delta=0$ the vortices of each component are organized in an Abrikosov lattice, but the vortices of different component are decoupled.
If $\delta$ is increased, the intercomponent interaction results in a repulsive force between the vortices of different components \cite{Eto:2011wp,Aftalion:2012} and the two component lattice has an hexagonal structure. 
When $\delta$ is increased further the unitary cell of the single component lattice is changed from a triangle to a square.
The value of $\delta$ for which the lattice reorganize depends on rotation speed.
If $\delta > 1$, phase separation occurs and vortex sheets appear \cite{Kasamatsu:2009}. 

We now want to systematically study the effect of an internal coherent coupling on the vortex lattice. 
When the two different components are atoms with different hyperfine spin states, the coherent coupling can be achieved by Rabi oscillations.
This results in the formation of a two vortex molecule, namely a dimer.

The potential induced by Rabi oscillations has the form
\begin{align}\label{eq:RabiPotential}
	V_R = - {\int d^{3}r}  \, \omega_R (\Psi_1^* \Psi_2 + \Psi_2^* \Psi_1) \, .
\end{align}
This interaction couples the atoms of the two components via the relative phase of the order parameters.
We expect the interplay between the Rabi interaction and the atom-atom interaction triggered by $g_{12}$ to be fundamental for vortex lattice structure. 
Then we consider the ratio $\omega_R/\delta$ as the relevant quantity.

\begin{figure}[t]
	\includegraphics[width=8cm]{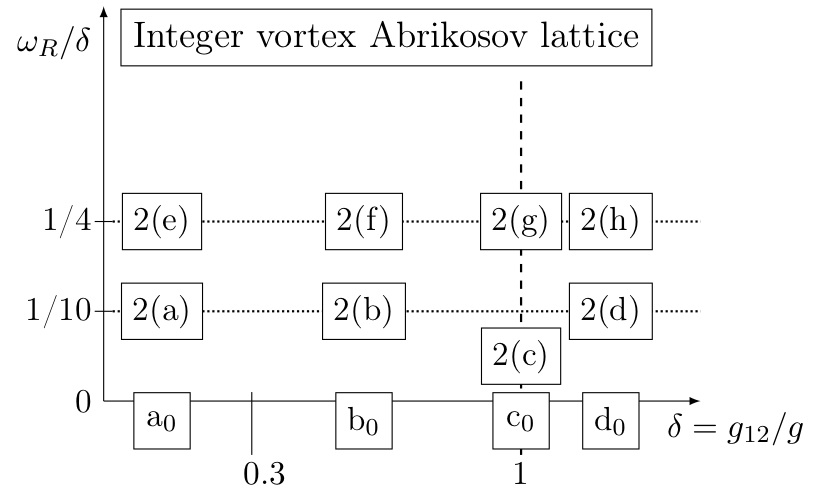}
	\caption{The vortex lattice phase diagram. $a_{0}$, $b_{0}$, $c_{0}$, $d_{0}$ indicate the lattices found in \cite{Kasamatsu:2003}. $a_{0}$ corresponds to the triangular lattice, $b_{0}$ to the square lattice, and $d_{0}$ to the vortex sheet, while $c_{0}$ represents the configuration staying at the boundary between lattice and vortex sheet regions.
	The others are the different configurations explained in the text. The labels refer to the pictures of Fig.~2.
	}
	\label{fig:RabiPhaseDiagram}
\end{figure}

\def\figwidth{3.1cm}

\begin{figure*}[ht]
\centering
\begin{tabular}{c@{\hskip 0.7cm}c}
	(a) $\delta= 0.2 , \omega_R= 0.03$ & (b) $\delta= 0.5 , \omega_R= 0.075$ \\ 
	\includegraphics[height=\figwidth]{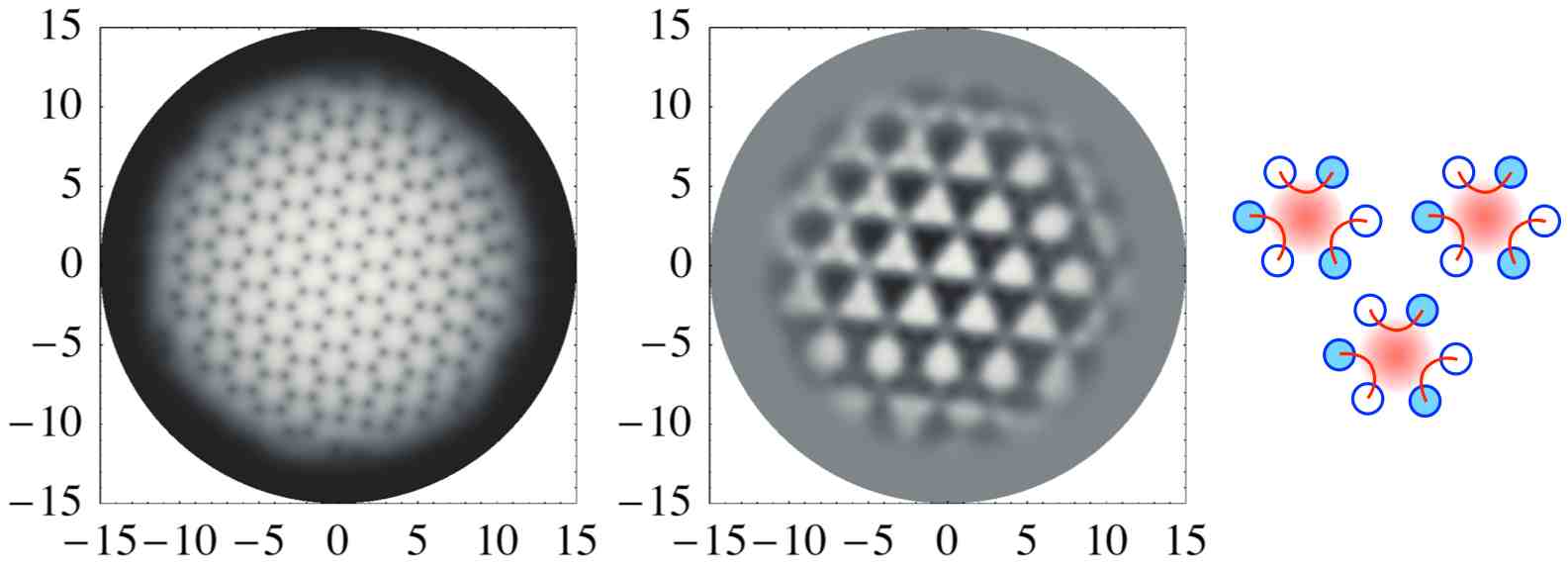} & \includegraphics[height=\figwidth]{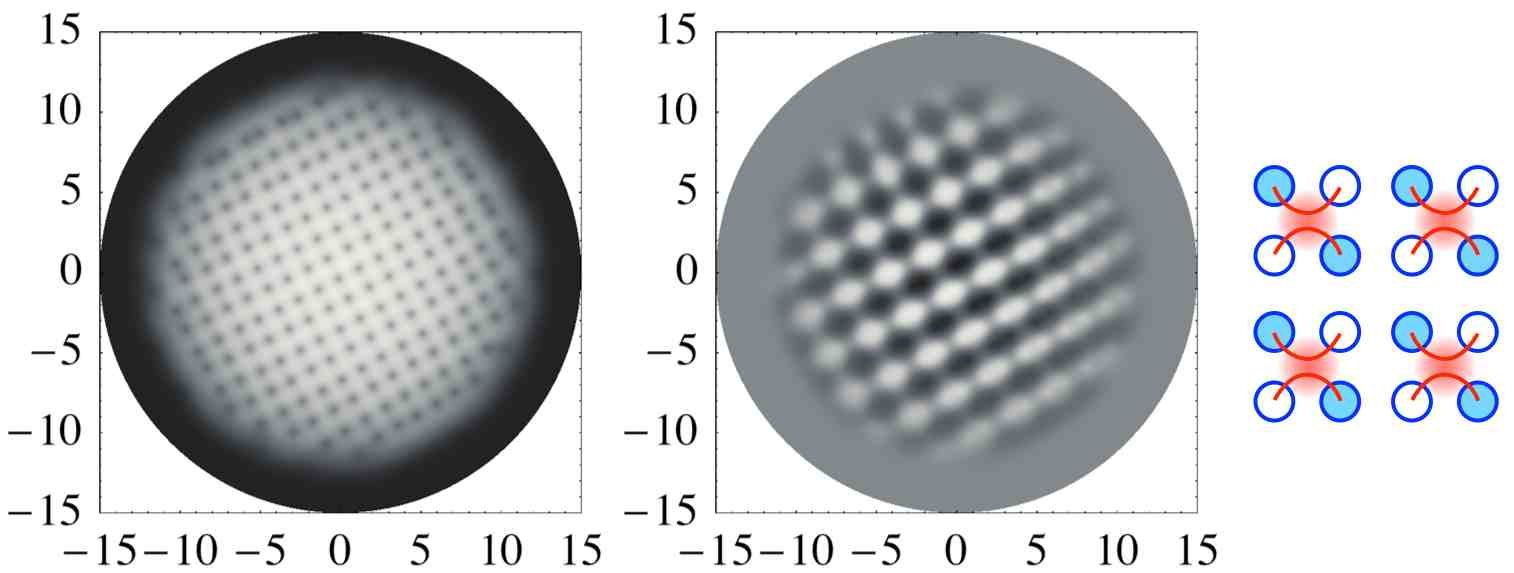} \\
	(c) $\delta= 1 , \omega_R= 0.05$ & (d) $\delta= 1.1 , \omega_R= 0.11$ \\
	\includegraphics[height=\figwidth]{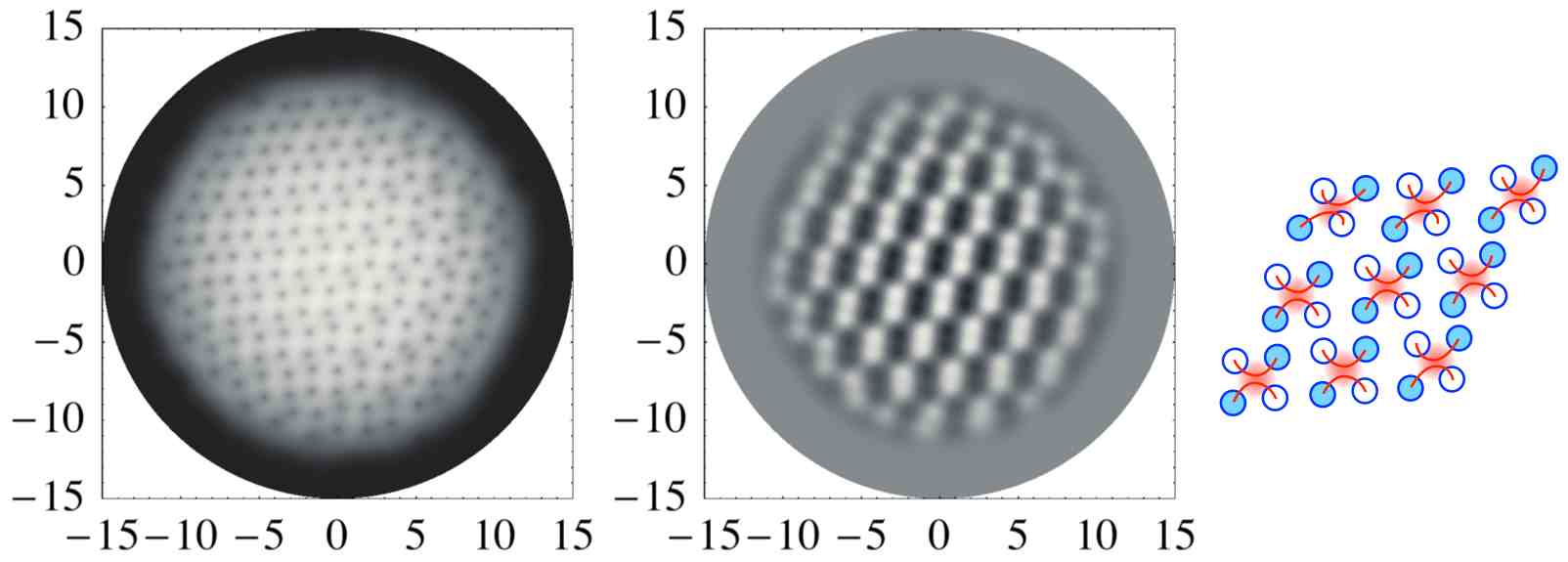} & \includegraphics[height=\figwidth]{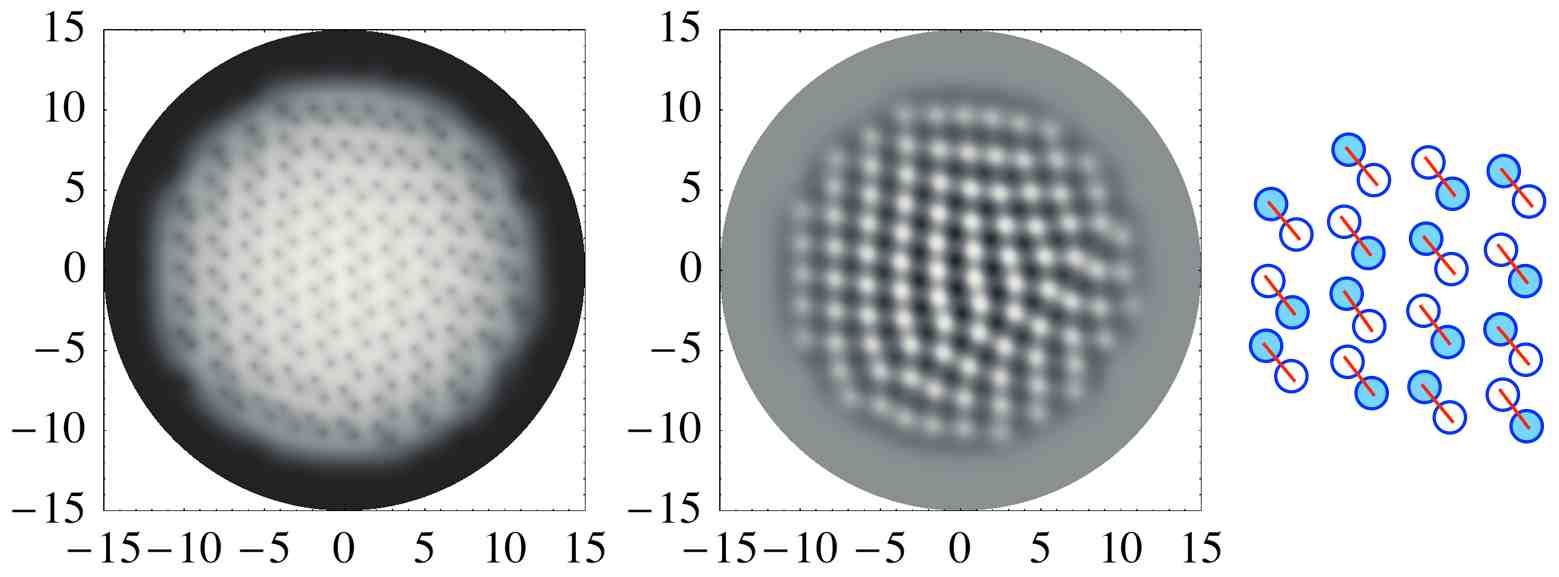} \\
	(e) $\delta= 0.2 , \omega_R= 0.05$ & (f) $\delta= 0.5 , \omega_R= 0.125$ \\
	\includegraphics[height=\figwidth]{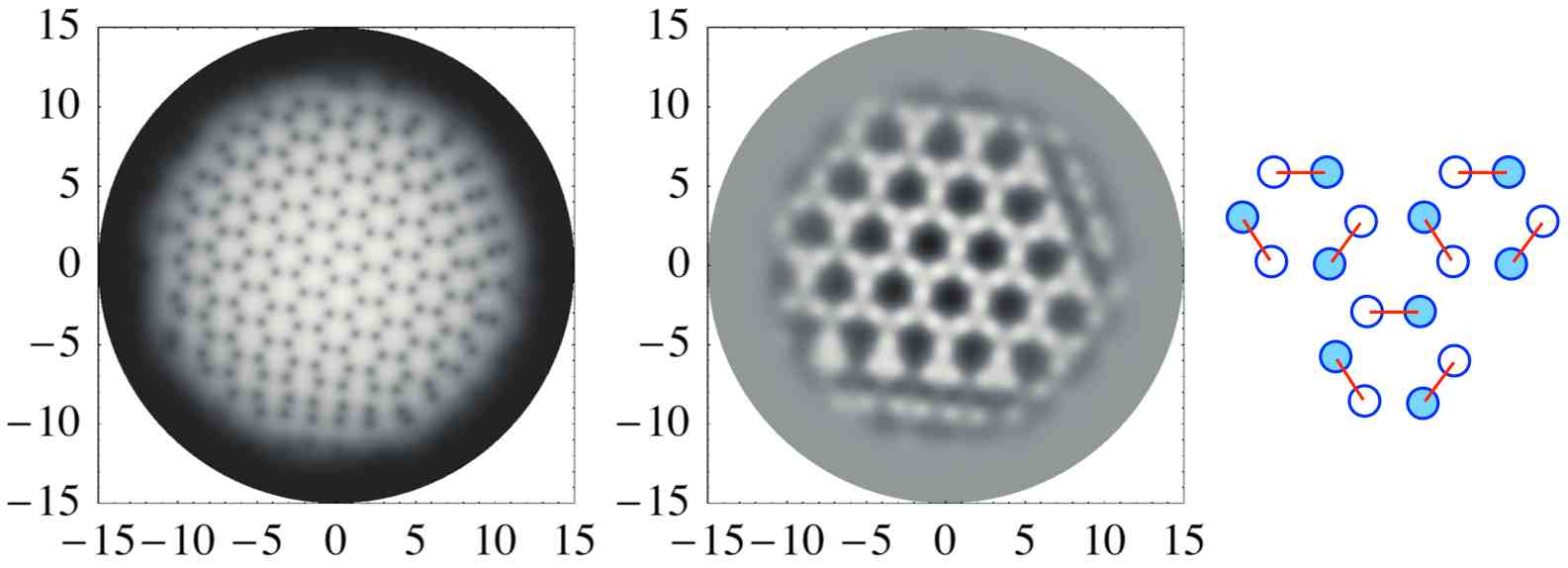} & \includegraphics[height=\figwidth]{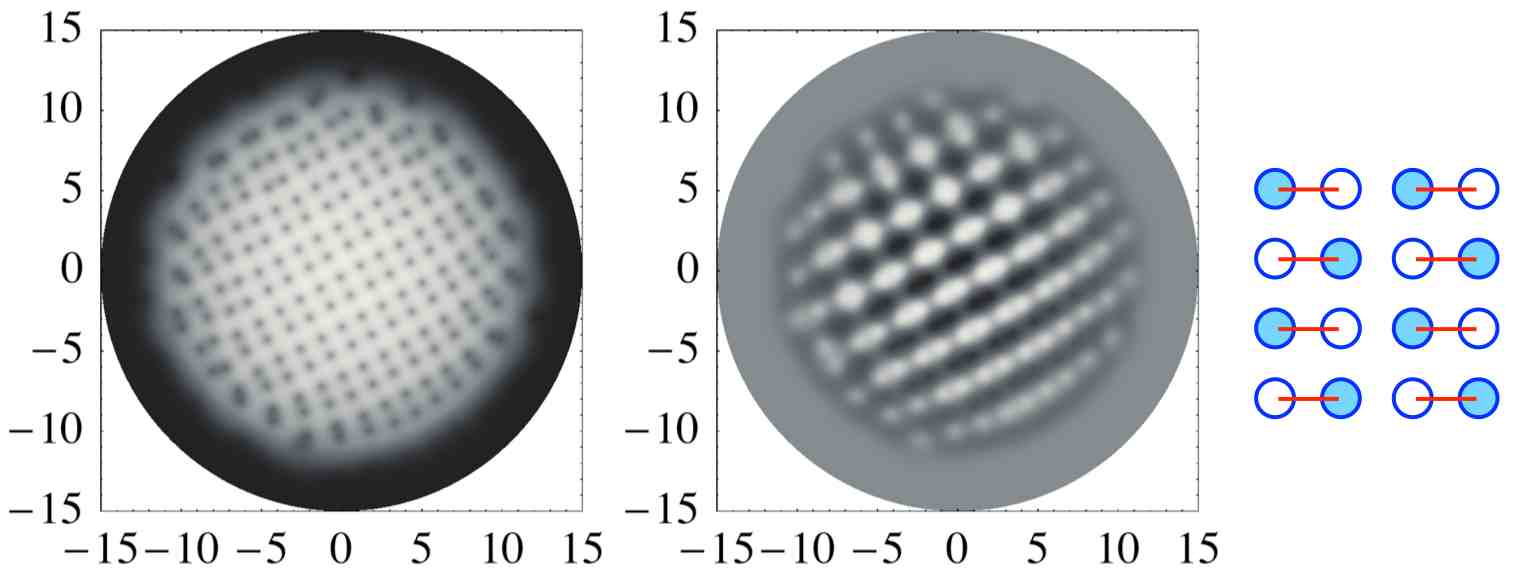} \\
	(g) $\delta= 1 , \omega_R= 0.25$ & (h) $\delta= 1.1 , \omega_R= 0.275$ \\
	\includegraphics[height=\figwidth]{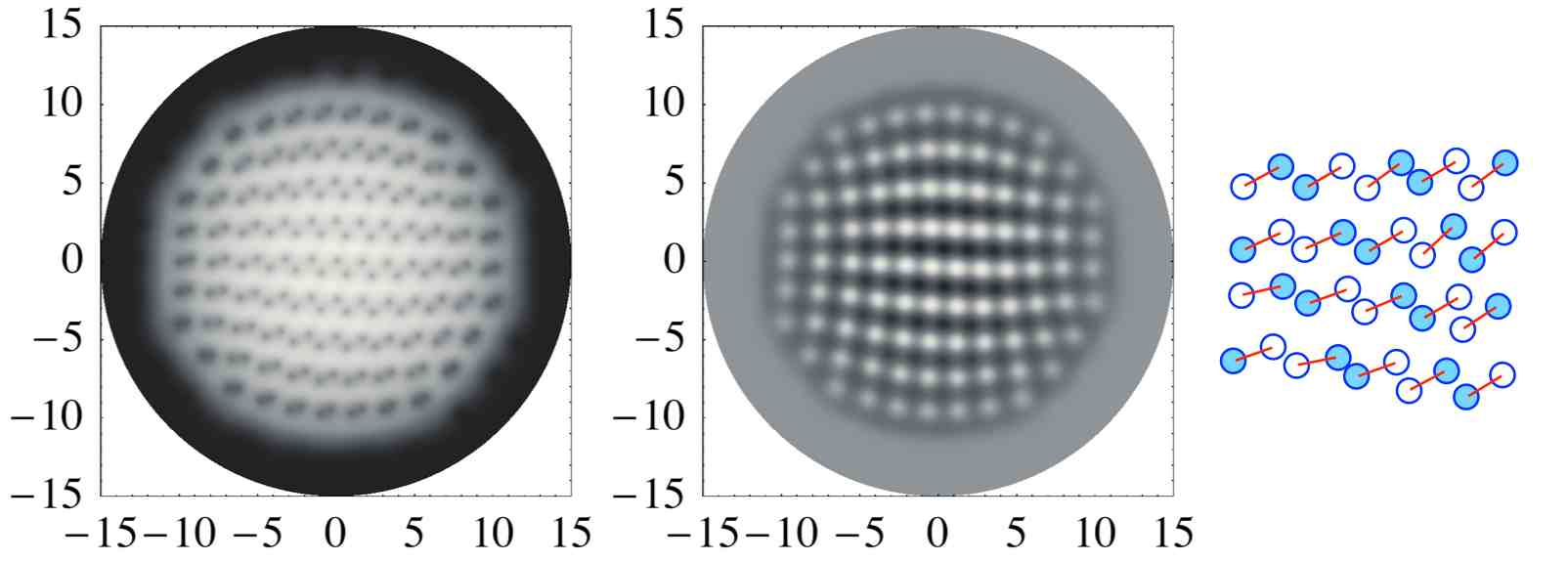} & \includegraphics[height=\figwidth]{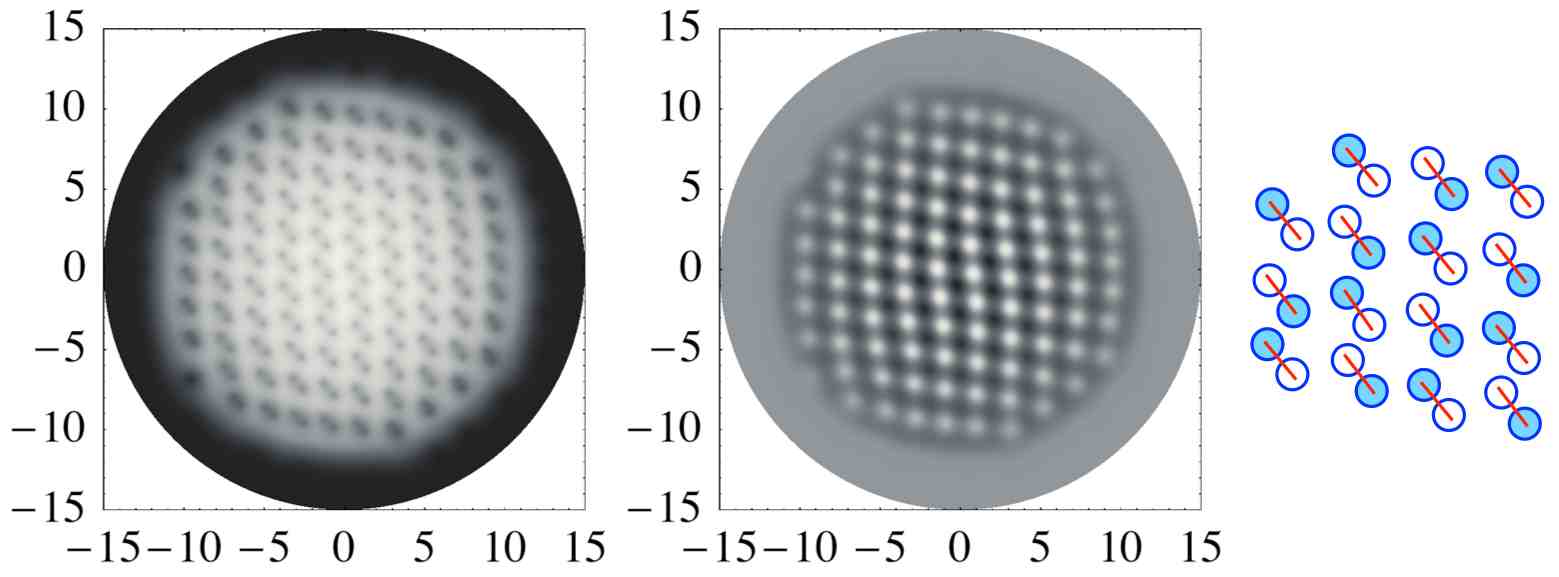} 
\end{tabular}
\caption{
In all subfigures the left panel is a plot of the density profile of the condensate, $n=|\Psi_{1}|^{2}+|\Psi_{2}|^{2}$ {(dark grey dots are vortices 
of the first or the second component)}, the middle panel is a plot of the Rabi energy in Eq.~\eqref{eq:RabiPotential} (white is positive, identified with domain walls), and the right panel is a schematic drawing of the lattice structure {The domain walls joining vortices are depicted as red lines and the vortices in the first and in the second component are distinguished by empty or filled circles}. 
The values of the parameters $\delta$ and $\omega_R$ are shown for each case.
The lattice defect in (d) appears just by chance
and in fact it resolves with increasing $\omega_R$, 
as can be seen in subfigure (h).
Panels (e-g) show the same cases of (a-c) with higher $\omega_R$. 
To visualize the domain walls, 
the plots of the phase difference between the two condensates are given in
the Appendix.
}
\label{fig:Fig2}
\end{figure*}
%%%%%%%%%%%%%%%%%%%%%%%%%%

\def\figwidth{8.6cm}

%%%%%%%%%%%%%%%%%%%%%%%%%%%%%%%%%%%%%
\begin{figure}[ht]
\centering
\begin{tabular}{cc} \vspace{-0.03cm}
(a)\;$\delta= 0.2 , \omega_R= 0.06$ \\ \vspace{-0.25cm}
\includegraphics[width=\figwidth]{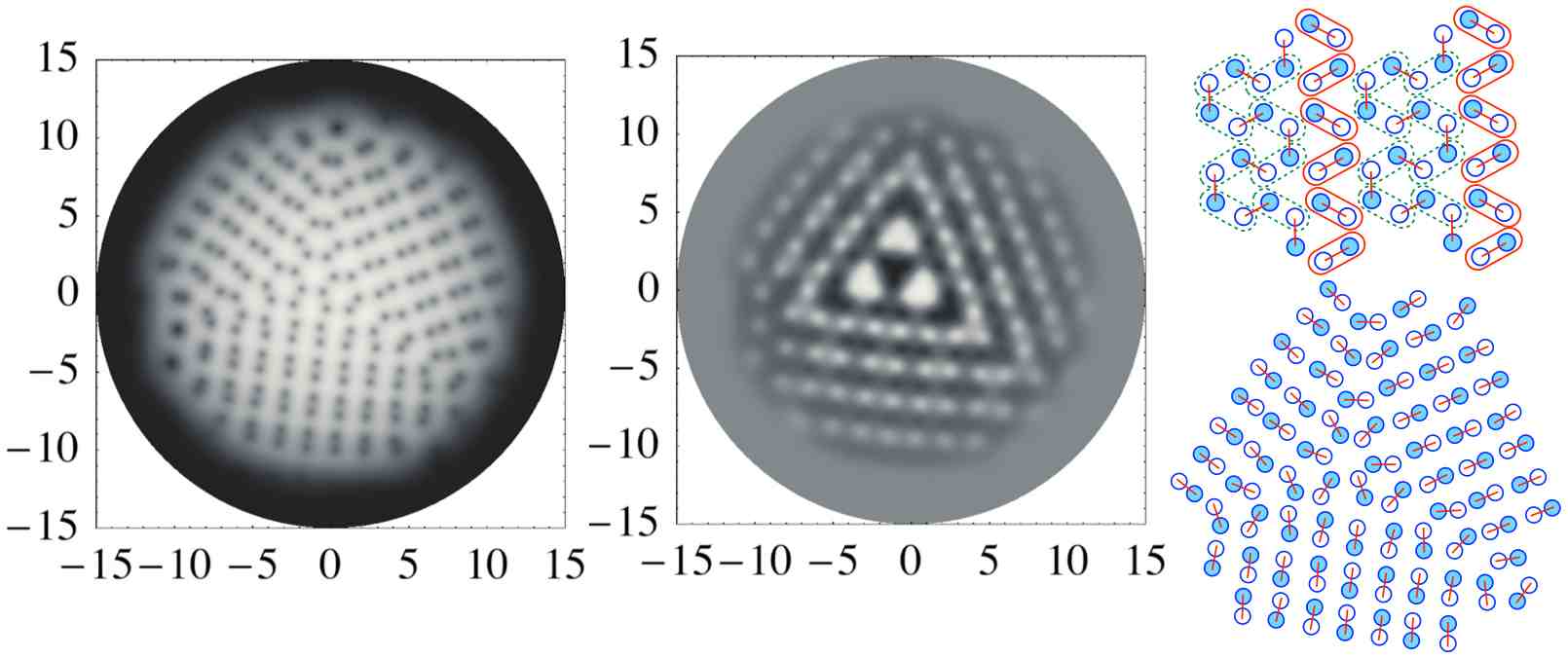} \\ 
(b)\;$\delta= 0.5 , \omega_R= 0.15$ \\ \vspace{-0.1cm}
\includegraphics[width=\figwidth]{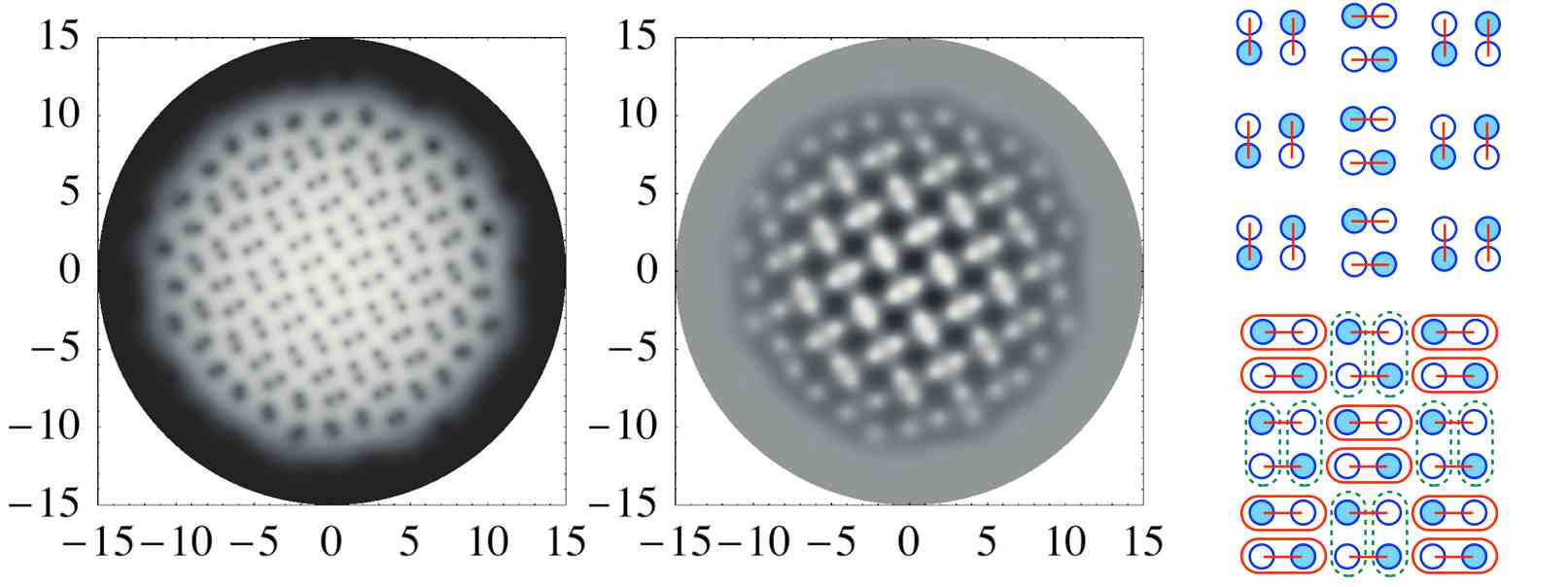} \\
(c)\;$\delta= 0.9 , \omega_R= 0.315$ \\ 
\includegraphics[width=\figwidth]{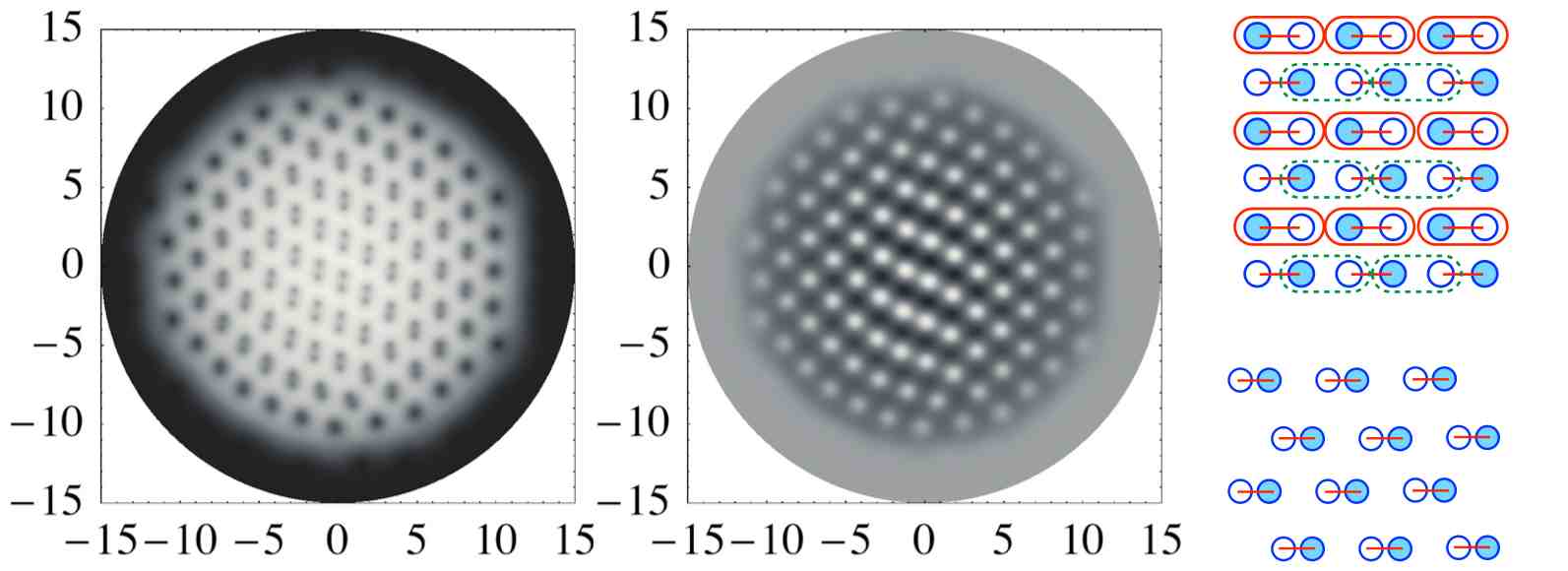} 
\end{tabular}
\caption{ (a-c)
The different partner changing patterns that we have identified. 
The left panel in each subfigure shows the plot of the density profile of the condensate 
$n=|\Psi_{1}|^{2}+|\Psi_{2}|^{2}$ {(dark grey dots are vortices 
of the first or the second component)} and the middle panel shows the Rabi energy in Eq.~\eqref{eq:RabiPotential} (white is positive, identified as domain walls). 
Schematic reproductions of the modifications on the dimers and of the lattice itself after the partner changing took place are shown in the right panel in each row. 
 {The domain walls joining vortices are depicted as red lines and the vortices in the first or in the second component are distinguished by empty or filled circles}. 
The domain walls are broken and reconnected to be new sets of dimers indicated by the dashed round boxes; the dimers indicated by solid round boxes are instead kept unchanged.
To visualize the domain walls, 
the plots of the phase difference between the two condensates are given 
in the Appendix.
}
\label{fig:Fig3}
\end{figure}

We first reproduced the results of \cite{Kasamatsu:2003} by numerically minimizing the free energy \eqref{eq:FreeEnergy}.
They are reported on the horizontal axis of Fig.~\ref{fig:RabiPhaseDiagram}.
We minimize the free energy~\eqref{eq:FreeEnergy} by 
the non-linear conjugate gradient method 
(the imaginary time propagation) 
in the FreeFem++ package. 
We calculate the ground state of the system with $\omega_R > 0$ by minimizing the energy functional \eqref{eq:FreeEnergy} with \eqref{eq:RabiPotential} added. 
We use the converged lattice solution obtained with $\omega_R=0$ as the starting-point configuration and then we increase the Rabi frequency by steps $\Delta\omega_R = \delta/20$. 
We then take the results of the converged calculation obtained for some value of $\omega_R$ as initial configuration for the next step.
By using this ``adiabatic'' evolution method we are confident that we always obtain the true ground state for the vortex lattice.
{We cross-checked our results by starting from the converged solution with $\delta=0$ and moving in the phase diagram along horizontal lines, with steps of $\Delta\delta = 0.1$, keeping the ratio $\omega_{R}/\delta$ fixed.
We report the results in the Appendix.}
%All the configurations in Figure~\ref{fig:RabiPhaseDiagram} proved to be true ground states.}
In all the numerical simulations we take $g=1$, $\Omega=0.98$ and $\mu_{1}=\mu_{2}={4.5}$.
{The number of atoms is $N_{1} = N_{2} = N \sim 10^{3}$, while the value of the healing length $\xi \sim 0.3 \, b_{\text{ho}}$.}
The ground state configurations obtained are schematically reported in Fig.~\ref{fig:RabiPhaseDiagram}. 
For simplicity we reported labels for the various configurations which we explain below.

We find that for $\omega_R/\delta\lesssim1/10$ multi-dimer bound states appear. 
When $\delta$ takes values corresponding to triangular lattices $0 < \delta \lesssim 0.3$, the system exhibits vortex hexamers connected by domain walls, as shown in  Fig.~\ref{fig:Fig2}(a); if $\delta$ is within the square lattice range $0.3 \lesssim \delta \lesssim~1$, each bound state is composed of four vortices making up tetramers, as shown in Fig.~2(b); when $\delta=1$ slanted tetramers appear, as shown in Fig.~2(c). 
This behavior is signaled by the form of the Rabi energy, whose maxima lie between a group of vortices. 
When the tension of neighboring domain wall and anti-domain wall is small enough they can bend toward each other and constitute a bound state, 
because of the attraction between them.
The structure of the lattice is schematically reported in the rightmost panels of Figs.~\ref{fig:Fig2} (a), \ref{fig:Fig2}(b) and \ref{fig:Fig2}(c).
However, these bound states are metastable and when ${1/10\lesssim \omega_R/\delta \lesssim 1/4}$ they split into dimers, as in Figs.~\ref{fig:Fig2}(e), \ref{fig:Fig2}(f), 
and 2(g).
The phase separation for $\delta>1$ prevents from the formation of multi-dimer bound states and only single dimers are formed even for higher values of $\omega_R/\delta$, 
as shown in Fig.~\ref{fig:Fig2}(d) and \ref{fig:Fig2}(h). 
This happens because, for $\delta>1$, even with very small $\omega_R$, there are no dimers whose domain walls are so close to bend to form multi-dimer molecules.
This can be seen from Fig.~\ref{fig:Fig2}(d) and \ref{fig:Fig2}(h).

When $\omega_R$ increases further, the lattice modifies drastically. 
The reason for this modification is that for large values of $\omega_R$ 
the dimers are undistinguishable from integer vortices, in which the positions of the vortices of {the first and the second} components coincide. 
Integer vortices repel each other and are organized in an Abrikosov lattice. 
However to reach the Abrikosov configuration, the various lattices of fractional vortices must be deeply modified.
This modification is achieved when $1/4\lesssim\omega_R/\delta\lesssim1/3$. 
The vortices change their partners. 

We were able to find the elementary patterns of rearrangement{, shown in Fig.~\ref{fig:Fig3}}. 
In the rightmost panel of each subfigure schematic representations of the process of partner changing are reported.
We indicate with round solid boxes the dimers left unchanged, while the dashed boxes represent the formation of new dimers.
When ${0<\delta\lesssim0.3}$ the lattice is triangular for ${\omega_R/\delta\lesssim 1/4}$, but when $\omega_R$ is increased the domain walls connecting vortices in dimers can break and reconnect as depicted in Fig.~\ref{fig:Fig3}(a). 
The partner changing process can be realized along three different directions.
This is due to the discrete rotational symmetry of the two-component hexagonal vortex lattice appearing when $\delta < 0.3$.
{All these possibilities are realized in general. When the lattice structure is very symmetric, as in the case of Fig.~\ref{fig:Fig3}(a), the lattice is divided into three domains} 
separated by a domain wall junction, 
as can be seen from the picture in the {left} panel where the density profile of the condensate is shown.
{In general, configurations with more defects are obtained, as shown in the Appendix.}

If ${0.3 \lesssim \delta < 1}$, the vortices can change partner in two different ways,
as shown in Fig.~\ref{fig:Fig3}(b) and \ref{fig:Fig3}(c). 
The rightmost panels represent the patterns of the partner changing processes, with the same conventions of Fig.~\ref{fig:Fig3}(a). 
As can be seen from the schematic diagram in Fig.~\ref{fig:Fig3}(b), four vortices forming a square cell change their partner, while links in the adjacent square cell are kept unchanged.
The situation is different in Fig.~\ref{fig:Fig3}(c), where vortices change their partner in alternating rows. 
The disposition of the dimers depicted in the bottom rightmost panel of Fig.~\ref{fig:Fig3}(c) is similar to that of Fig.~\ref{fig:Fig2}(h).
{However,
%However the relative ordering of the vortices belonging to different components is different.
for $\delta>1$ the free energy is minimized when vortices of the same component are close to each other, forming vortex sheets. }
Instead in the case of Fig.~\ref{fig:Fig3}(c), the lower energy is achieved when vortices belonging to the same component are well separated. 
{This explains why in \ref{fig:Fig2}(h) vortices of the same component of adjacent dimers are close to each other, while in \ref{fig:Fig3}(c) they are not.}
{The rearrangement of Figure~\ref{fig:Fig3}(b) is difficult to obtain in the whole lattice; nevertheless, limited portions of the lattice always show this pattern, in the indicated parameter ranges, as reported in the Appendix.
The pattern of Figure~\ref{fig:Fig3}(c) is frequently obtained when $\delta$ is in the square lattice range.}

In conclusion we have found a rich variety of lattices which can appear in two-component BECs 
where internal coherent coupling is induced.
We introduced the interaction via Rabi oscillations and studied systematically its effect on the vortex lattice, increasing gradually the frequency $\omega_R$ from $\omega_R=0$. 
The vortex lattice is reorganized when $\omega_R$ is increased up to high values in order to reach the triangular lattice configuration where all dimers become integer vortices. 
During this process multi-dimer bound states are formed and vortices can exchange partners in different ways, depending on the form of the fractional vortex lattice. 
The patterns of the exotic vortex lattices 
which we have found in this Letter 
can be easily distinguished from usual 
triangular or square lattice from observations 
for instance by the time of flight.

\section{Acknowledgements} 
M.~C. thanks the Department of Physics at Hiyoshi, Keio University, for warm hospitality in the beginning of this project.
M.~N. thanks INFN, Pisa, for partial support and hospitality while this work was done.
Both authors thank Kenichi Kasamatsu, Yasumoto Tanaka, Walter Vinci and Kenichi Konishi for useful discussions.
The numerical calculations were performed on the INFN CSN4 cluster located in Pisa.
The work of M.~N. is supported in part by 
KAKENHI  (No. 23740198, No. 23103513, No. 25400268 and No. 25103720).

\section{Appendix}

Here we present more results about the domain walls connecting the vortices when coherent coupling is induced between two components.
We want to show more specifically the bending which the domain walls exhibit when the Rabi coupling $\omega_R$ is small and multi-dimer bound states are formed and also clarify more the partner changing schemes.
If we take the wave functions of the two component with the form
\begin{equation}
	\Psi_{1}({\bf r}) = |\Psi_{1}({\bf r})| e^{i\theta_{1}({\bf r})} \, , \quad \Psi_{2}({\bf r}) = |\Psi_{2}({\bf r})| e^{i\theta_{2}({\bf r})} \, ,
\end{equation} 
we can define the phase difference between the two components as 
\begin{equation}
	\phi({\bf r}) = \theta_{1}({\bf r}) - \theta_{2}({\bf r}) \, .
\end{equation}
We can then write the Rabi energy term as
\begin{align}\label{eq:AppRabiEnergy}
	V_R &= - \omega_R (\Psi_1^*({\bf r}) \Psi_2 ({\bf r}) + \Psi_2^*({\bf r}) \Psi_1 ({\bf r})) = -2 \, \omega_R \, |\Psi_{1}({\bf r})| |\Psi_{2}({\bf r})| \cos (\phi({\bf r})) \, . 
\end{align}
To decrease this term, $\phi({\bf r})$ must be close to zero modulo $2\pi$, with the amplitudes of the wave functions increased in the overlapping region.
Then sine-Gordon domain walls form between vortices.

In our calculation we see that this domain wall can bend and then multi-dimer bound states can form in the vortex lattice, when the Rabi coupling $\omega_R$ is small enough.
In Fig.~\ref{fig:PhaseDiffBound} we show the plots of the phase difference $\phi({\bf r})$ for some set of the parameters $\{\delta=g_{12}/g,\omega_R\}$.
Fig.~\ref{fig:PhaseDiffBound}(a)--(h) correspond to Fig.2(a)--(h) in 
the main text, respectively.
In the top panels we plot \eqref{eq:AppRabiEnergy} where the whiter regions are maxima. 
The middle panels are plots of the phase difference, in which domain walls between vortices are clearly visible.
In the bottom panels we superimpose the phase difference plot with the plot of the condensate density $n=|\Psi_{1}|^{2}+|\Psi_{2}|^{2}$, where the light gray dots at the extrema of domain walls are vortices belonging to different components.
In the first three columns we see that domain walls are bent and multi-dimer molecules are formed. 
In the case of Fig.~\ref{fig:PhaseDiffBound}(a) we find hexamers, while tetramers appear in Fig.~\ref{fig:PhaseDiffBound}(b).
In Fig.~\ref{fig:PhaseDiffBound}(c) domain walls are less bent than in the two previous cases because the elementary cell of the lattice is a slanted square, so that it is more difficult for them to interact with each other.
In all the three cases, we notice from the Rabi energy plot that the domain walls interact and bind together three or two dimers.
In  Fig.~\ref{fig:PhaseDiffBound}(d) domain walls are not bent at all, even with very small Rabi coupling. 
This is due to the fact that the dimers are disposed in an alternating pattern, so that no domain wall is faced with the other.

In Fig.~\ref{fig:PhaseDiffBound} (e)--(h) we report the same plots obtained with the same values for $\delta$, but with higher $\omega_R$.
We can see that the domain walls are now mostly straight and all the bound states split into single dimers.

In Fig.~\ref{fig:PartnerChanging} we show the analogous plots for a higher particular value for $\omega_R$, at which vortices change their partner with which they form the dimers, as explained in the main text. 
Fig.~\ref{fig:PartnerChanging}(a)--(c) correspond to Fig.~3(a)--(c) in 
the main text, respectively.
From this plots it is clear the disposition of the dimers after the partner changing process.
In Fig.~\ref{fig:PartnerChanging}(a) we see that some domain walls are bent and multi-dimer bound states are possible in the center of the cloud, where a domain wall junction is created between the three different domains in which the lattice is divided.
Fig.~\ref{fig:PartnerChanging}(b) shows that the dimers are now formed with different orientation, compared with Fig.~\ref{fig:PhaseDiffBound}(f); Fig.~\ref{fig:PartnerChanging}(c) presents a different pattern of partner changing, realized when $\delta$ is close to one.

{As explained in the main text, the results have been confirmed by starting from the converged solution with $\delta=0$ and moving in the phase diagram along horizontal lines, with steps of $\Delta\delta = 0.1$, keeping the ratio $\omega_{R}/\delta$ fixed.
By following this method, we found the configurations reported in Fig.~3 and Fig.~4, which can be compared to the ones of Fig.~2 in the main text.
Therefore, all the configurations in Fig.~2 in the main text prove to be ground states, and can be all included in the phase diagram.}

{Regarding the configurations in Fig.~3(a)-(b) in the text, it is quite difficult to find the same regular configurations in the whole lattice; nevertheless, they are always present in some limited regions.
We can conclude that the patterns described in Fig.~3(a)-(b) are always realized locally and sometimes globally.
On the other hand, Fig.~3(c) in the main text is always realized when $\delta \gtrsim 0.9$.}

\def\figwidth{3.5cm}

\begin{figure*}[ht]
\centering
\begin{tabular}{c@{\hskip 0.4cm}c@{\hskip 0.4cm}c@{\hskip 0.4cm}c}
	(a) $\delta= 0.2 , \omega_R= 0.03$ & (b) $\delta= 0.5 , \omega_R= 0.075$ & (c) $\delta= 1 , \omega_R= 0.05$ & (d) $\delta= 1.1 , \omega_R= 0.11$ \\
	\includegraphics[width=\figwidth]{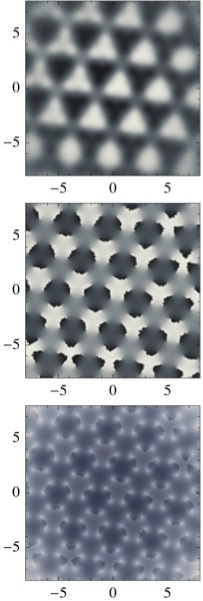} & \includegraphics[width=\figwidth]{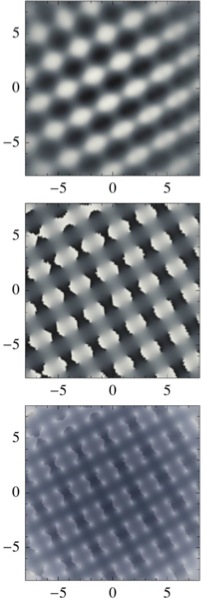} & \includegraphics[width=\figwidth]{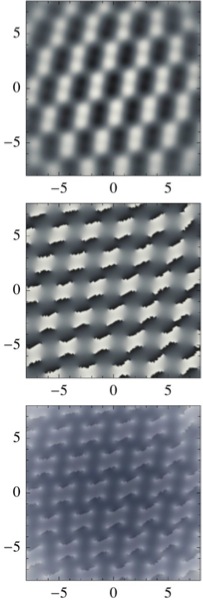} & \includegraphics[width=\figwidth]{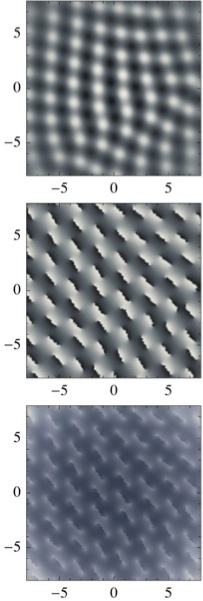} \\
%%%%%%%%%%%
%%%%%%%%%%%
	(e) $\delta= 0.2 , \omega_R= 0.03$ & (f) $\delta= 0.5 , \omega_R= 0.075$ & (g) $\delta= 1 , \omega_R= 0.05$ & (h) $\delta= 1.1 , \omega_R= 0.11$ \\
	\includegraphics[width=\figwidth]{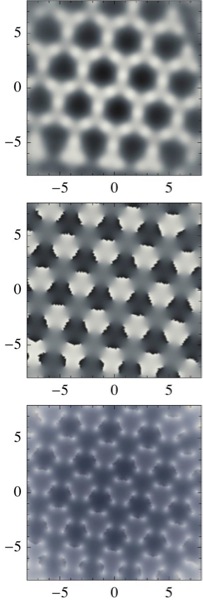} & \includegraphics[width=\figwidth]{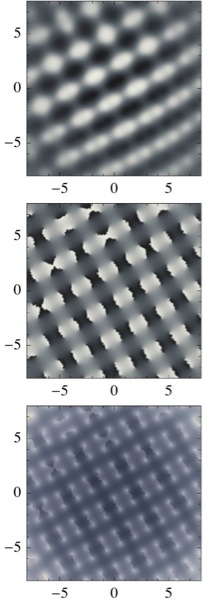} & \includegraphics[width=\figwidth]{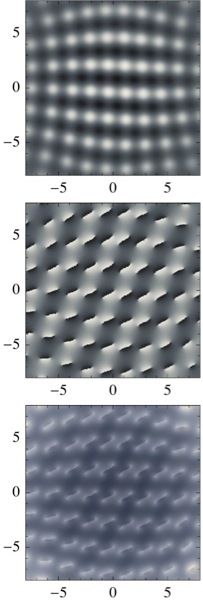} & \includegraphics[width=\figwidth]{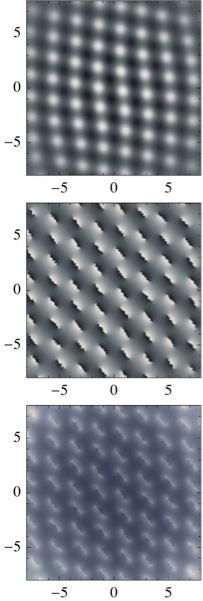} \\
\end{tabular}
\caption{
In all subfigures the top panel is a plot of the Rabi energy in Eq.~\eqref{eq:AppRabiEnergy} (white is positive) {and} the middle panel is a plot of the phase difference $\phi({\bf r})$,{ where white corresponds to $\phi(\textbf{r})=\pi$ and black to $\phi(\textbf{r})=-\pi$. The domain wall is placed at $\phi(\textbf{r})=\pm \pi$ and can be identified with boundary between white and black regions. The bottom panel is a superposition of the middle panel with the plot of the condensate density $n=|\Psi_{1}|^{2}+|\Psi_{2}|^{2}$, in which colors have been inverted with respect to the left panel of Fig. 2 in the main text to make this plot more easily readable; consequently here vortices are identified with light gray dots.} 
The values of the parameters $\delta$ and $\omega_R$ are shown for each case.
In (a) a lattice of hexamers is visible, while in (b) we find tetramers.  
The tetramers are slanted in (c) with a different value of $\delta$. 
No multi-bound states of dimers appears for $\delta>1$, in (d).
In (a), (b) and (c) domain walls are bent and multi-dimers bound states are formed.
In (e), (f) and (g) the bound states are mostly split into single dimers.  
In (h) we see that the lattice is periodic with straight domain walls.
}
\label{fig:PhaseDiffBound}
\end{figure*}
%%%%%%%%%%%%%%%%%%%%%%%%%%

\begin{figure*}[ht]
\centering
\begin{tabular}{ccc}
(a)\;$\delta= 0.2 , \omega_R= 0.06$ & 
(b)\;$\delta= 0.5 , \omega_R= 0.15$ & 
(c)\;$\delta= 0.9 , \omega_R= 0.36$ \\
	\includegraphics[width=\figwidth]{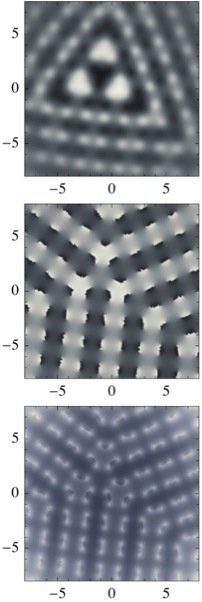} & \includegraphics[width=\figwidth]{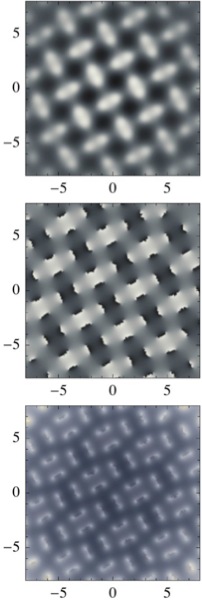} & \includegraphics[width=\figwidth]{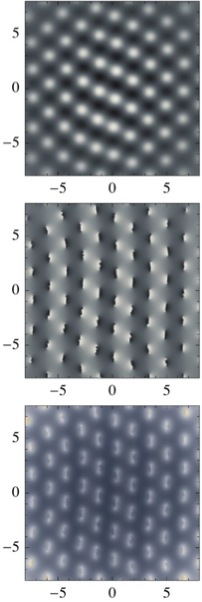}  \\

\end{tabular}
\caption{
The different partner changing patterns that we have identified. 
In all subfigures, the top panel is a plot of the Rabi energy in Eq.~\eqref{eq:AppRabiEnergy} (white is positive), {and} the middle panel is a plot of the phase difference $\phi({\bf r})$,{ where white corresponds to $\phi(\textbf{r})=\pi$ and black to $\phi(\textbf{r})=-\pi$. The domain wall is placed at $\phi(\textbf{r})=\pm \pi$ and can be identified with boundary between white and black regions. The bottom panel is a superposition of the middle panel with the plot of the condensate density $n=|\Psi_{1}|^{2}+|\Psi_{2}|^{2}$, in which colors have been inverted with respect to the left panel of Fig. 2 in the main text to make this plot more easily readable; consequently here vortices are identified with light gray dots.} 
In (a) we show the partner changing process for the hexagonal lattice, appearing when $\delta < 0.3$.
The defect in (a) is physically relevant, representing a junction between different domains.
The subfigures (b) and (c) represent the two partner-changing patterns realized in square vortex lattice, ${0.3\lesssim\delta<1}$.
In these two cases no defect appeared in the lattice, which exhibits a periodic structure.
}
\label{fig:PartnerChanging}
\end{figure*}

\def\figwidth{12cm}

\begin{figure*}
\centering
\begin{tabular}{c}
	(a) $\delta = 0.2 , \, \omega_{R} = 0.02 , \, \omega_{R}/\delta = 0.1$ \\
	\includegraphics[width=\figwidth]{{{GridPlot_G12_0.2_Gamma_0.02}}} \\
	(b) $\delta = 0.5 , \, \omega_{R} = 0.05 , \, \omega_{R}/\delta = 0.1$ \\
	\includegraphics[width=\figwidth]{{{GridPlot_G12_0.5_Gamma_0.05}}}\\
	(c) $\delta = 1 , \, \omega_{R} = 0.1 , \, \omega_{R}/\delta = 0.1$ \\
	\includegraphics[width=\figwidth]{{{GridPlot_G12_1_Gamma_0.1}}}\\
	(d) $\delta = 1.1 , \, \omega_{R} = 0.11 , \, \omega_{R}/\delta = 0.1$ \\
	\includegraphics[width=\figwidth]{{{GridPlot_G12_1.1_Gamma_0.11}}}
\end{tabular}
\caption{{The configurations obtained by keeping the ratio $\omega_{R}/\delta$ fixed and gradually increasing $\delta$.
In all subfigures, the left panel shows the plot of the density $n=|\Psi_{1}|^{2}+ |\Psi_{2}|^{2}$, {where light gray dots are identified with vortices}; the middle panel is a plot of the Rabi energy Eq. \eqref{eq:AppRabiEnergy} (white regions correspond to positive values); and the right panel is the plot of the phase difference $\phi(\textbf{r})$, where white corresponds to $\phi(\textbf{r})=\pi$ and black to $\phi(\textbf{r})=-\pi$. The domain wall is placed at $\phi(\textbf{r})=\pm \pi$ and can be identified with boundary between white and black regions. The corresponding parameters are shown above all subfigures.
All the subfigures (a)-(d) are comparable with Fig.~2(a)-(d) reported in the main text, proving that these configurations are reliable ground states.}}
\end{figure*}

\begin{figure*}
\centering
\begin{tabular}{c}
	(a) $\delta = 0.2 , \, \omega_{R} = 0.05 , \, \omega_{R}/\delta = 0.25$ \\
	\includegraphics[width=\figwidth]{{{GridPlot_G12_0.2_Gamma_0.05}}} \\
	(b) $\delta = 0.5 , \, \omega_{R} = 0.125 , \, \omega_{R}/\delta = 0.25$ \\
	\includegraphics[width=\figwidth]{{{GridPlot_G12_0.5_Gamma_0.125}}}\\
	(c) $\delta = 1 , \, \omega_{R} = 0.25 , \, \omega_{R}/\delta = 0.25$ \\
	\includegraphics[width=\figwidth]{{{GridPlot_G12_1_Gamma_0.25}}}\\
	(d) $\delta = 1.1 , \, \omega_{R} = 0.275 , \, \omega_{R}/\delta = 0.25$ \\
	\includegraphics[width=\figwidth]{{{GridPlot_G12_1.1_Gamma_0.275}}}
\end{tabular}
\caption{{The configurations obtained by keeping the ratio $\omega_{R}/\delta$ fixed and gradually increasing $\delta$.
In all subfigures, the left panel shows the plot of the density $n=|\Psi_{1}|^{2}+ |\Psi_{2}|^{2}$, where light gray dots are identified with vortices; the middle panel is a plot of the Rabi energy Eq. \eqref{eq:AppRabiEnergy} (white regions correspond to positive values); and the right panel is the plot of the phase difference $\phi(\textbf{r})$, where white corresponds to $\phi(\textbf{r})=\pi$ and black to $\phi(\textbf{r})=-\pi$. The domain wall is placed at $\phi(\textbf{r})=\pm \pi$ and can be identified with boundary between white and black regions.
The corresponding parameters are shown above all subfigures.
All the subfigures (a)-(d) are comparable with Fig.~2(e)-(h) reported in the main text, proving that these configurations are reliable ground states.
The defects in (c) and (d) appear by chance.}}
\end{figure*}

\begin{figure*}
\centering
\begin{tabular}{c}
	(a) $\delta = 0.2 , \, \omega_{R} = 0.06 , \, \omega_{R}/\delta = 0.3$ \\
	\includegraphics[width=\figwidth]{{{GridPlot_G12_0.2_Gamma_0.06}}} \\
	(b) $\delta = 0.5 , \, \omega_{R} = 0.15 , \, \omega_{R}/\delta = 0.3$ \\
	\includegraphics[width=\figwidth]{{{GridPlot_G12_0.5_Gamma_0.15}}}\\
	(c) $\delta = 0.9 , \, \omega_{R} = 0.315 , \, \omega_{R}/\delta = 0.35$ \\
	\includegraphics[width=\figwidth]{{{GridPlot_G12_0.9_Gamma_0.315}}}
\end{tabular}
\caption{{The configurations obtained by keeping the ratio $\omega_{R}/\delta$ fixed and gradually increasing $\delta$.
In all subfigures, the left panel shows the plot of the density $n=|\Psi_{1}|^{2}+ |\Psi_{2}|^{2}$, where light gray dots are identified with vortices; the middle panel is a plot of the Rabi energy Eq. \eqref{eq:AppRabiEnergy} (white regions correspond to positive values); and the right panel is the plot of the phase difference $\phi(\textbf{r})$, where white corresponds to $\phi(\textbf{r})=\pi$ and black to $\phi(\textbf{r})=-\pi$. The domain wall is placed at $\phi(\textbf{r})=\pm \pi$ and can be identified with boundary between white and black regions.
The corresponding parameters are shown above all subfigures.
In (a) and (b) the two patterns of partner changing revealed in Fig.~3(a), 3(b) in the main text are realized in limited portions of the lattice. 
On the other hand, the pattern of Fig.~3(c) in the main text is observed in the whole lattice when $0.9 \lesssim \delta < 1 $.}}
\end{figure*}

\end{document}